\documentclass[a4paper]{article}
\usepackage{INTERSPEECH2022}
\usepackage{amsmath,graphicx}
\usepackage{multirow}
\usepackage{booktabs}
\usepackage{color,makecell,url,times, tabularx,bbm,amssymb,multirow,rotating,diagbox}

\title{Improving Target Sound Extraction with Timestamp Information}
\name{Helin Wang$^{1 \dagger}$\thanks{$^{\dagger}$ Indicates equal contribution.}, 
Dongchao Yang$^{1 \dagger}$, Chao Weng$^{2}$, Jianwei Yu$^{2}$, Yuexian Zou$^{1,*}$\thanks{$^{*}$ Corresponding Author.}}
\address{$^1$ADSPLAB, School of ECE, Peking University, Shenzhen, China\\
  $^2$Tencent AI Lab, Shenzhen, China}
\email{zouyx@pku.edu.cn}

\begin{document}

\maketitle
\begin{abstract}
  Target sound extraction (TSE) aims to extract the sound part of a target sound event class from a mixture audio with multiple sound events.
  The previous works mainly focus on the problems of weakly-labelled data, jointly learning and new classes, however, no one cares about the onset and offset times of the target sound event, which has been emphasized in the auditory scene analysis. In this paper, we study to utilize such timestamp information to help extract the target sound via a target sound detection network and a target-weighted time-frequency loss function.
  More specifically, we use the detection result of a target sound detection (TSD) network as the additional information to guide the learning of target sound extraction network. We also find that the result of TSE can further improve the performance of the TSD network, so that a mutual learning framework of the target sound detection and extraction is proposed. In addition, a target-weighted time-frequency loss function is designed to pay more attention to the temporal regions of the target sound during training. Experimental results on the synthesized data generated from the Freesound Datasets show that our proposed method can significantly improve the performance of TSE.
\end{abstract}

\section{Introduction}

A wide variety of sounds exist in the world, 
which provide critical information about our surroundings,
such as the violin sound in a concert and klaxons on the street.
Our daily lives can be greatly improved if we could develop hearing devices that have the ability of selecting the sound we are interested in
and removing the sound we are not interested in.
Such target sound extraction and removal applications have been studied these years in machine hearing,
aiming at target speaker \cite{vzmolikova2019speakerbeam,wang2019voicefilter,lin2021sparsely}, 
music instruments \cite{slizovskaia2021conditioned,LeeCL19},
and sound events \cite{borsdorf2021universal,delcroix2021few,gfeller2021one,pishdadian2020learning,kong2020source}.
In this paper, 
we focus on the target sound extraction problem, 
which is particularly challenging due to the large variety of sounds it covers 
(\textit{e.g.} baby cry, telephone call, animal sounds, etc).

Target sound extraction (TSE) consists of extracting the sound of a target sound event class from a mixture audio,
which uses an extraction network that estimates the target sound given the sound mixture 
and an embedding vector that represents the characteristics of the target sound. 
The embedding vector can be obtained using an embedding encoder that receives either an enrollment audio sample \cite{gfeller2021one} or a 1-hot vector that represents the target sound event class \cite{kong2020source,ochiai2020listen}. 
The extraction neural network and the embedding encoders can be jointly trained.
Several researchers further explored the problems of new classes and weakly-labelled data.
For example, Delcroix \textit{et al.} \cite{delcroix2021few} combined 1-hot-based and enrollment-based target sound extraction to improve the performance of new classes.
Pishdadian \textit{et al.} \cite{pishdadian2020learning} proposed a method to solve the weakly-supervised sound separation with a sound classification network,
and Gfeller \textit{et al.} \cite{gfeller2021one} solved this problem by randomly mixing two sounds and using one as both the reference audio and the target sound.
Although these previous works show good performance,
no one cares about the timestamp information of the target sound,
\textit{i.e.} the onset and offset times.
Such information is quite important in the auditory scene analysis,
which has been widely studied like sound event detection \cite{cakir2017convolutional,parascandolo2016recurrent} and voice activity detection \cite{sohn1999statistical,zhang2012deep},
and can be explored to help guide the localization and extraction of target sound.

\begin{figure}[t]
  \centering
  \includegraphics[width=1.0\linewidth]{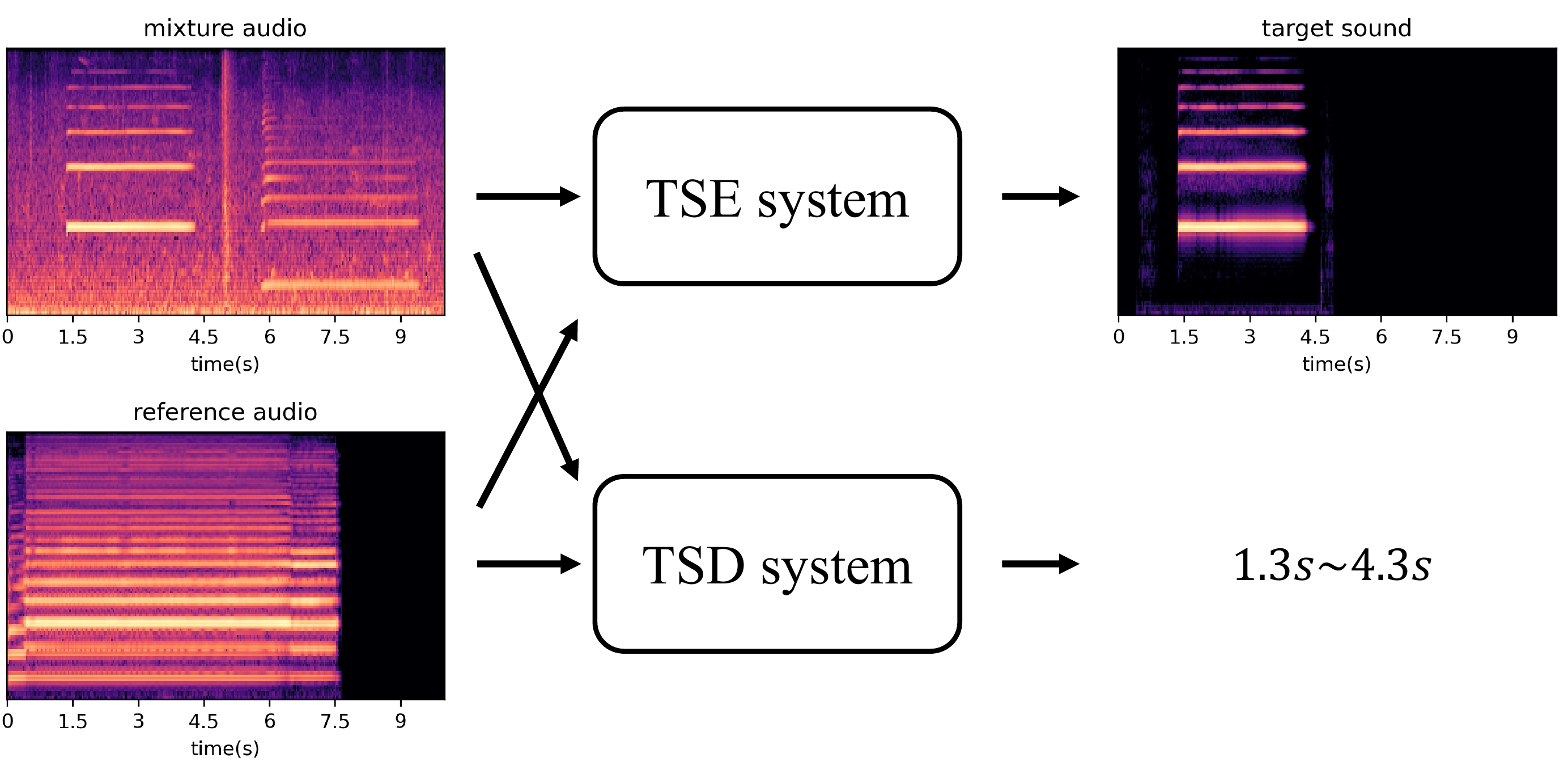}
  \caption{Example of TSE and TSD tasks. Both input the mixture audio and a reference audio. TSE aims to directly output the target sound while TSD aims to output the timestamps that the target sound happen within the mixture audio.}
  \label{fig:a}
  \vspace*{-\baselineskip}
\end{figure}

In this paper, we explicitly study to utilize the timestamp information for TSE,
and propose to use an additional detection network along with a target-weighted time-frequency loss function.
To be more specific,
we can obtain the result of target sound detection (TSD) \cite{yang2021detect} given the mixture audio and the embedding vector,
and this result is applied as the additional timestamp information.
Then, the TSE network inputs the mixture audio, the embedding vector and the detection result,
and outputs the extracted result.
Furthermore,
we find that the result of TSE can also help TSD,
by using it as another input audio reference.
In this way,
a mutual learning framework is proposed to jointly improve the performance of target sound extraction and detection.
In addition,
in order to enhance the model's ability to deal with the overlapping parts of the target sound and others,
we combine the time-domain and frequency-domain loss function and giving more weights to the overlapping timestamp when calculating the loss function,
which is called a target-weighted time-frequency loss function.
Experiments are carried out on the simulated data generated from sound events taken from the Freesound Datasets (FSDs) \cite{fonseca2017freesound,fonseca2018general}.
Our proposed method can obtain an overall SI-SDR improvement of $11.81$ dB and a SI-SDR improvement of $6.14$ dB for the temporal regions in which the target sound happens, which provides $15.2\%$ and $19.5\%$ improvement over our baseline method, respectively.


\begin{figure}[t]
  \centering
  \includegraphics[width=1.0\linewidth]{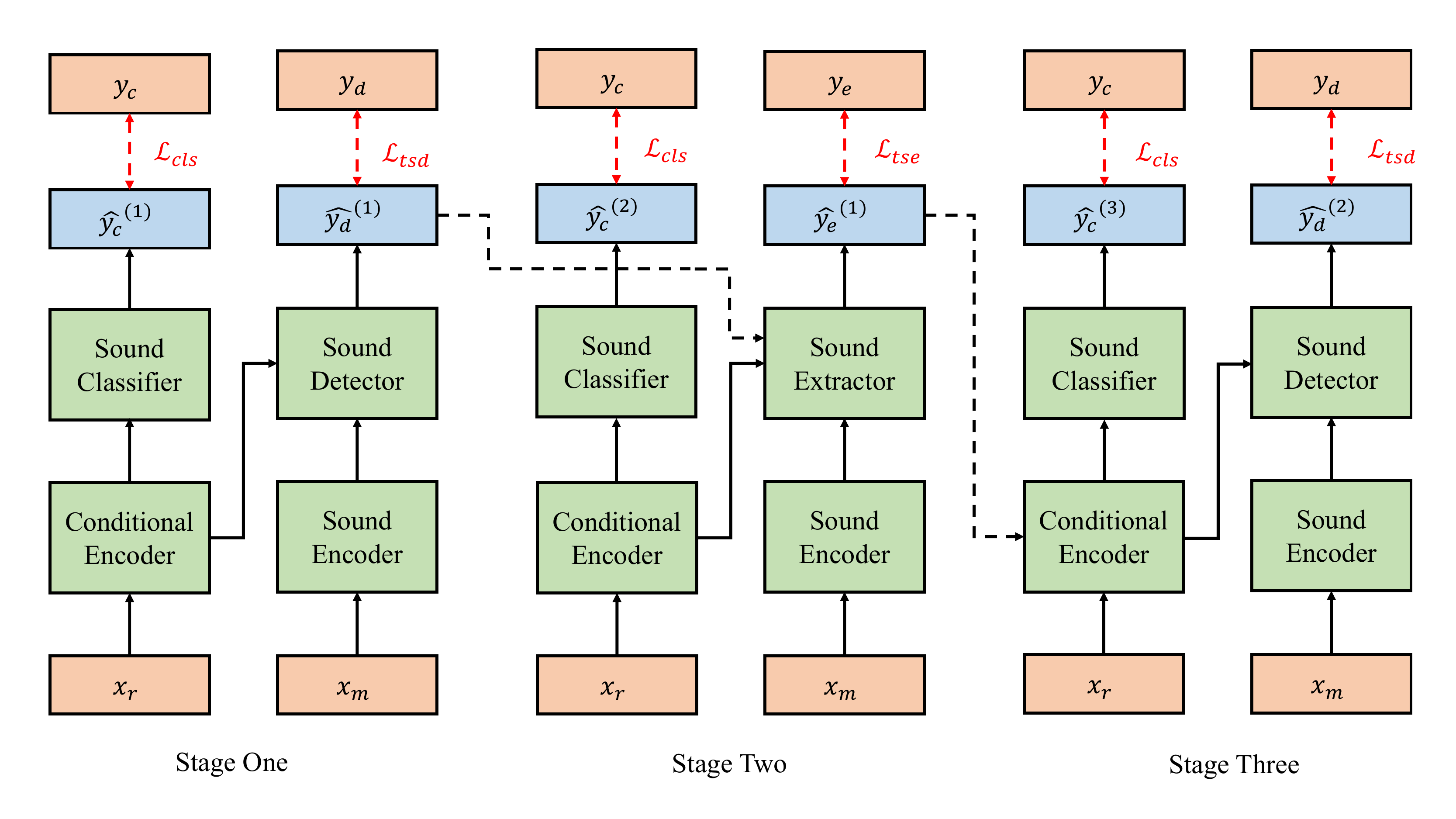}
  \caption{Overall architecture of Tim-TSENet. In Stage One, we train a target sound detection network which outputs the predicted detection result. 
  In Stage Two, we train a target sound extraction network using the predicted detection result from Stage One.
  Then in Stage Three, the target sound detection network is updated using the extracted result from Stage Two.}
  \label{fig:1}
  \vspace*{-\baselineskip}
\end{figure}

\section{Proposed Method}
In this section, 
we will introduce our proposed method that uses the timestamp information for the target sound extraction, which is called Tim-TSENet.
The architecture of Tim-TSENet is shown in Figure~2,
which shows three stages for the training of target sound extraction and detection networks.
Let’s begin with a brief review of the baseline system. 
We then introduce the loss function and the mutual learning framework.

\subsection{Target sound extraction and detection}
Figure~1 shows an example of TSE and TSD tasks.
Given a reference audio that represents the target sound event, TSE aims to extract the target sound from a mixture audio (a regression task),
while TSD aims to get the detection result of the target sound (a classification task).
Both need a conditional encoder to obtain the embedding vector of the reference audio and a sound encoder to extract acoustic features from the mixture audio.
The difference is that TSE needs a sound extractor to output the estimated sound but TSD needs a sound detector to output the predicted scores.
More specifically,
given the reference audio $\boldsymbol{x_r} \in \mathcal{R}^{N_{r}}$ and the mixture audio $\boldsymbol{x_m} \in \mathcal{R}^{N_{m}}$,
a conditional encoder $f_{con}(\cdot)$ is applied to get the embedding vector $\boldsymbol{e} \in \mathcal{R}^{N_{e}}$.
\begin{equation}
    \label{eq1}
    \boldsymbol{e} = f_{con}(\boldsymbol{x_r};\theta_{con})
\end{equation}
where $\theta_{con}$ denotes the parameters of the conditional encoder. $N_{r}$, $N_{m}$ and $N_{e}$ are the dimension of the reference audio, the mixture audio and the embedding vector respectively. Then a sound encoder $f_{enc}(\cdot)$ is applied to get the acoustic features $\boldsymbol{a} \in \mathcal{R}^{N_{t} \times N_{f}}$.
\begin{equation}
    \label{eq2}
    \boldsymbol{a} = f_{enc}(\boldsymbol{x_m};\theta_{enc})
\end{equation}
where $\theta_{enc}$ denotes the parameters of the sound encoder, $N_{t}$ denotes the number of frames for the acoustic features and $N_{f}$ denotes the dimension of each frame. For the TSE task, 
a sound extractor $f_{ext}(\cdot)$ is finally employed to obtain the estimated target sound $\boldsymbol{\hat{y_{e}}} \in \mathcal{R}^{N_{m}}$.
\begin{equation}
    \label{eq3}
    \boldsymbol{\hat{y_{e}}} = f_{ext}(\boldsymbol{a}, \boldsymbol{e};\theta_{ext})
\end{equation}
Here, $\theta_{ext}$ denotes the parameters of the sound extractor.
Note that the dimension of $\boldsymbol{\hat{y_{e}}}$ is the same as $\boldsymbol{y_m}$.
For the TSD task, 
a sound detector $f_{det}(\cdot)$ is applied to obtain the predicted scores of all the frames $\boldsymbol{\hat{y_{d}}} \in \mathcal{R}^{N_{t}}$.
\begin{equation}
    \label{eq4}
    \boldsymbol{\hat{y_{d}}} = f_{det}(\boldsymbol{a}, \boldsymbol{e};\theta_{det})
\end{equation}
where $\theta_{det}$ denotes the parameters of the sound detector.

\subsection{Loss function}
For the sound separation tasks (\textit{e.g.} speech separation \cite{wang2018supervised, kavalerov2019universal,huang2015joint,isik2016single} and speech enhancement \cite{xu2013experimental,benesty2006speech}), there are two popular loss functions.
One is the frequency-domain loss function,
which starts with calculating the short-time Fourier transform (STFT) to create a time-frequency (T-F) representation of the mixture audio.
The T-F bins that correspond to the target sound are then separated, and are used to
synthesize the source waveform using inverse STFT.
In this case, the loss function is formulated by the mean square error (MSE) between the T-F bins estimated target sound $\boldsymbol{\hat{s}} \in \mathcal{R}^{T \times F}$ and the corresponding ground truth $\boldsymbol{s} \in \mathcal{R}^{T \times F}$.
\begin{equation}
    \label{eq5}
    \mathcal{L}_{fmse} = \frac{1}{T \times F}\Vert \boldsymbol{s} - \boldsymbol{\hat{s}} \Vert^2
\end{equation}
where $\Vert \cdot \Vert^{2}$ denotes the $l_2$ norm, $T$ denotes the number of frames and $F$ denotes the number of frequency bins for each frame. Such frequency-domain is friendly to the computational complexity and shows good performance in many applications. 
However, the phase information is sometimes not totally considered \cite{wang2018supervised}, and suitable window size and hop size of STFT need to be designed.
Another way is calculating the loss directly on the time domain,
including the time-domain MSE loss and the time-domain scale-invariant source-to-distortion ratio (SI-SDR) loss \cite{luo2018tasnet,luo2019conv}.
Given the estimated waveform $\boldsymbol{\hat{y_{e}}} \in \mathcal{R}^{N_{m}}$ and the corresponding ground truth $\boldsymbol{y_{e}} \in \mathcal{R}^{N_{m}}$, 
the time-domain MSE loss is formulated by:
\begin{equation}
    \label{eq6}
    \mathcal{L}_{tmse} = \frac{1}{N_{m}}\Vert \boldsymbol{y_{e}} - \boldsymbol{\hat{y_{e}}} \Vert^2
\end{equation}
Here, $N_{m}$ denotes the dimension of the mixture audio. This time-domain MSE loss is sensitive to the variation of the energy of the waveform and can be used to limit the energy of the estimated waveform to the ground truth.
In addition, the SI-SDR loss is defined as:
\begin{equation}
\label{eq7}
\mathcal{L}_{sisdr}=10 \log _{10}\left(\frac{\left\|\frac{\langle\boldsymbol{\hat{y_{e}}}, \boldsymbol{y_{e}}\rangle}{\langle \boldsymbol{y_{e}}, \boldsymbol{y_{e}}\rangle} \boldsymbol{y_{e}}\right\|^{2}}{\left\|\frac{\langle\boldsymbol{\hat{y_{e}}}, \boldsymbol{y_{e}}\rangle}{\langle \boldsymbol{y_{e}}, \boldsymbol{y_{e}}\rangle} \boldsymbol{y_{e}}-\boldsymbol{\hat{y_{e}}}\right\|^{2}}\right)
\end{equation}
where $\langle\cdot, \cdot\rangle$ is the inner product. To ensure scale
invariance, the signals $\boldsymbol{\hat{y_{e}}}$ and $\boldsymbol{y_{e}}$ are normalized to zero-mean prior to the SI-SDR calculation. 
This loss is designed to directly optimize the evaluation metric (\textit{i.e.} source-to-distortion ratio \cite{chen2017deep}) and has widely used in speech separation.
Based on the complementary of these three loss functions, we propose to combine them, which forms a time-frequency loss functions for the TSE task.
\begin{equation}
    \label{eq8}
    \mathcal{L}_{tse} = \mathcal{L}_{fmse} + \alpha \mathcal{L}_{tmse} + \beta \mathcal{L}_{sisdr}
\end{equation}
where $\alpha$ and $\beta$ are the hyper-parameters to balance the loss function.
Such loss function treats the whole audio equally,
however, for a TSE task, we focus more on the temporal regions in which the target sound happens with in the whole audio clip.
In this paper,
in order to enhance the model's ability to deal with such regions,
we propose a target-weighted time-frequency loss function that gives more weights to the overlapping timestamps when calculating the loss.
Assume that the target sound happens in the timestamp $[N_{m_1}, N_{m_2}]$ of the mixture audio where $N_{m_2} \le N_{m}$, we can get the losses $\mathcal{L}_{fmse-t}$ (target-region frequency-domain MSE loss), $\mathcal{L}_{tmse-t}$ (target-region time-domain MSE loss) and $\mathcal{L}_{sisdr-t}$ (target-region time-domain SI-SDR loss) according to equations~(5-7).
The target-weighted time-frequency loss function is then calculated by:
\begin{align}
    \label{eq9}
    \mathcal{L}_{tse-t} &= \mathcal{L}_{fmse-t} + \alpha \mathcal{L}_{tmse-t} + \beta \mathcal{L}_{sisdr-t} \\
    \mathcal{L}_{tse-w} &= \mathcal{L}_{tse} + \tau \mathcal{L}_{tse-t}
\end{align}
Here, $\tau$ is another hyper-parameter to control the ratio of the weighted part.
For the TSD task,
given the predicted score $\boldsymbol{\hat{y_{d}}} \in \mathcal{R}^{N_{t}}$ and the frame-level ground truth label $\boldsymbol{y_{d}} \in \mathcal{R}^{N_{t}}$,
the TSD network can be optimized by minimize the binary cross-entropy (BCE) loss function:
\begin{equation}
    \label{eq10}
    \mathcal{L}_{tsd} = \sum_{i=1}^{N_t}(-y_d^i\log\hat{y_d}^i-(1-y_d^i)\log(1-\hat{y_d}^i))
\end{equation}
where $y_d^i \in \{0, 1\}$, $\hat{y_d}^i \in [0, 1]$ and $N_t$ denotes the number of frames.
Both TSD and TSE can be jointly optimized by a sound classification task to improve the quality of the embedding vector.
As shown in Figure~2, a sound classifier $f_{cls}(\cdot)$ is applied to get the predicted classification result $\boldsymbol{\hat{y_{c}}} \in \mathcal{R}^{C}$.
\begin{equation}
    \label{eq11}
    \boldsymbol{\hat{y_{c}}} = f_{cls}(\boldsymbol{e};\theta_{cls})
\end{equation}
where $\theta_{cls}$ denotes the parameters of the sound classifier and $C$ denotes the number of classes.
Given the clip-level ground truth label, $\boldsymbol{y_{c}} \in \mathcal{R}^{C}$,
the cross-entropy loss function \cite{wang2020environmental} is then employed.
\begin{equation}
    \label{eq12}
    \mathcal{L}_{cls} = \sum_{i=1}^{C}(-y_c^i\log\hat{y_c}^i-(1-y_c^i)\log(1-\hat{y_c}^i))
\end{equation}
Note that $y_c^i \in \{0, 1\}$ and $\hat{y_c}^i \in [0, 1]$. The final loss function for the TSE and TSD with jointly learning is:
\begin{align}
    \label{eq13}
    \mathcal{L}_{tse-j} &= \mathcal{L}_{tse-w} + \mathcal{L}_{cls} \\
    \mathcal{L}_{tsd-j} &= \mathcal{L}_{tsd} + \mathcal{L}_{cls}
\end{align}

\subsection{Mutual learning framework}
TSE and TSD tasks provide different views of the target sound within the mixture audio, where TSE focuses on the separation and TSD focuses on the timestamp information.
Motivated by this, we propose a mutual learning framework to improve the two tasks.
The TSE network can utilize the timestamp information from the TSD result to guide the separation of target sound, and the TSD network can use the extracted target sound from the TSE result as an additional reference audio to enhance the reference information.
Figure~2 shows an example of three-stage training architecture.
In Stage One,
a TSD network is trained with equation~(15) as the loss function.
Then in Stage Two, this TSD network is fixed and we then train the TSE network by loss function~(14).
Here, we use the result of the TSD network $\boldsymbol{\hat{y_d}}^{(1)}$ as the additional timestamp information,
and concatenate it with the embedding vector $\boldsymbol{e}$.
So the equation~(3) is modified by:
\begin{equation}
    \label{eq14}
    \boldsymbol{\hat{y_e}}^{(1)} = f_{ext}(\boldsymbol{a}, \boldsymbol{e}, \boldsymbol{\hat{y_d}}^{(1)};\theta_{ext})
\end{equation}
Note that we upsample or downsample $\boldsymbol{\hat{y_d}}^{(1)}$ to the same number of frames as $\boldsymbol{a}$.
Finally, for the Stage Three,
we fix the TSE network and update the TSD network. Except the original reference audio $\boldsymbol{x_r}$,
the result from the TSE network $\boldsymbol{\hat{y_e}}^{(1)}$ is also used in this stage.
We can get two embeddings from the two audios, and we use the average value as the new embedding.
\begin{align}
    \label{eq15}
    \boldsymbol{e}^{\prime} &= f_{con}(\boldsymbol{\hat{y_e}}^{(1)};\theta_{con}) \\
    \boldsymbol{\hat{y_{d}}}^{(2)} & = f_{det}(\boldsymbol{a}, \frac{\boldsymbol{e}+\boldsymbol{e}^{\prime}}{2};\theta_{det})
\end{align}
In this way, TSE and TSD can help each other and we argue that this training framework can be recycled to further improve the performance. In this work,
considering the cost of training time,
we only attempt three stages.

\section{Experiments}

\subsection{Datasets}
Following \cite{ochiai2020listen}, we create datasets of simulated sound event mixtures based on the Freesound Dataset Kaggle 2018 corpus (FSD) \cite{fonseca2017freesound,fonseca2018general}, which contains audio clips from 41 diverse sound event classes, such as human sounds, object sounds, musical instruments. The duration of the audio samples ranges from 0.3 to 30 seconds. We generate 10 seconds mixtures by randomly choosing one target sound and 1-3 interference sound clips from the FSD corpus and adding them to random time-positions on top of the 10 seconds background sound. We choose the background sound from the acoustic scene classification task of Detection and Classification of Acoustic Scenes and Events (DCASE) 2019 Challenge \cite{mesaros2018multi}. Furthermore, the signal-noise ratio (snr) is set randomly between -5 and 10 dB for each foreground sound. If the duration of the foreground sound is longer than 10 seconds, we only keep the first 10 seconds of it. In addition, we found that some foreground sounds contain silent zones, and we use a pre-processing algorithm \cite{wang2019affects} to remove these zones. We also record the onset and offset times of the target sound according to the time-positions and the duration information. The onset and offset information can then be used to train the target sound detection task. Note that the onset and offset information is only used in the training stage. For each mixture audio, we randomly choose a reference audio that has the same class with the target sound. We also pad or cut the reference audio to 10 seconds.
In the experiments, we down-sample the audios to 16 kHz to reduce the computational and memory costs. The training, validation and test sets consist of 47356, 16000 and 16000 samples respectively.

\begin{table}[t] \centering
\caption{Performance of different models. 
Here, 'f', 't', 's' and 'w' represents the frequency-domain MSE loss, the time-domain MSE loss,
the time-domain SI-SDR loss and the target-weighted time-frequency loss respectively. All the experiments of Tim-TSENet use the jointly learning of TSE and a classification task. Note that the results of \cite{delcroix2021few} and \cite{ochiai2020listen} are reproduced under the experimental setups with ours.}
\label{tab:1}
\begin{tabular}{|c|p{0.22cm}p{0.22cm}p{0.22cm}p{0.22cm}|c|c|}
\hline
\multirow{2}{*}{Model} & \multicolumn{4}{|c|}{Loss function} & \multirow{2}{*}{SI-SDRi} & \multirow{2}{*}{SI-SDRi-t} \\
\cline{2-5}
& f & t & s & w & & \\
\hline
\multirow{13}{*}{\textbf{Tim-TSENet}} & $\surd$ & $\times$ & $\times$ & $\times$ & 7.61 & 2.90  \\
& $\times$ & $\surd$ & $\times$ & $\times$ & 4.39  & 2.31 \\
& $\times$ & $\times$ & $\surd$ & $\times$ & 3.63  & 1.86  \\
& $\surd$ & $\surd$ & $\times$ & $\times$ & 7.44  & 3.14  \\
& $\surd$ & $\times$ & $\surd$ & $\times$ & 10.22  & 5.06 \\
& $\times$ & $\surd$ & $\surd$ & $\times$ &9.22  & 4.68 \\
& $\surd$ & $\surd$ & $\surd$ & $\times$ & 10.25 & 5.14 \\
\textbf{(ours)}& $\surd$ & $\surd$ & $\surd$ & 0.2 & 10.88  & 5.65 \\
& $\surd$ & $\surd$ & $\surd$ & 0.5 & 10.68  & 5.52  \\
& $\surd$ & $\surd$ & $\surd$ & 1.0 & 10.77 & 5.65 \\
& $\surd$ & $\surd$ & $\surd$ & 1.2 & 10.86  & 5.77 \\
& $\surd$ & $\surd$ & $\surd$ & 1.5 & \textbf{10.91} & \textbf{5.83} \\
& $\surd$ & $\surd$ & $\surd$ & 1.8 & 10.87 & 5.76 \\
& $\surd$ & $\surd$ & $\surd$ & 2.0 & 10.79  & 5.63  \\
\hline
\cite{delcroix2021few} &  \multicolumn{4}{|c|}{$\times$} & 9.87 & 4.69 \\
\cite{ochiai2020listen} &  \multicolumn{4}{|c|}{$\times$} & 10.02 & 4.74 \\
\hline
\end{tabular}
\end{table}

\subsection{Metrics}
For the TSE task, we use the mostly-used scale-invariant signal-to-distortion ratio improvement (SI-SDRi).
In addition, in order to evaluate the model's ability to extract the target sound in the temporal regions where it happens, we apply a novel metric called scale-invariant signal-to-distortion ratio improvement in the target sound regions (SI-SDRi-t).
For the TSD task, the segment-based F-measure (segment-F1) and event-based F-measure (event-F1) \cite{yang2021detect} are employed.

\subsection{Implementation details}
For the sound extractor in Tim-TSENet, we use the Conv-tasnet with the original settings \cite{luo2019conv} due to its effectiveness and efficiency.
For the sound detector, we use the same CRNN structure as \cite{yang2021detect}.
Besides, following \cite{yang2021detect}, a four-block convolutional network \cite{kong2020panns} pretrained on Audioset \cite{gemmeke2017audio} is applied as the conditional encoder for both TSE and TSD.
The dimension of the embedding vector is $128$.
We use the Adam optimizer \cite{kingma2015adam} with an initial learning rate of $1 \times 10^{-3}$ and a weight decay of $1 \times 10^{-5}$.
The whole training epochs are $60$ for TSE and $100$ for TSD.
We carry out the experiments three times and report the mean value.

\subsection{Experimental results}
Table~1 reports the performance of our proposed Tim-TSENet and two state-of-the-art methods \cite{delcroix2021few,ochiai2020listen}.
We firstly test Tim-TSENet with different settings of loss function,
\textit{i.e.} the frequency-domain MSE loss, the time-domain MSE loss,
the time-domain SI-SDR loss and the target-weighted time-frequency loss.
For a single one among the first three loss, we can see that the time-domain SI-SDR loss preforms the worst. An interesting thing is that for the speech-related tasks \cite{luo2019conv}, the time-domain SI-SDR loss is always the state-of-the-art method. We argue that the separation of speech and sound event has a obvious gap because of different sampling rate, different frequency distribution and different duration.
By combining the time-domain and frequency-domain loss, we can get our baseline performance of 10.25 for SI-SDRi and 5.14 for SI-SDRi-t.
Comparing with \cite{delcroix2021few} and \cite{ochiai2020listen}, our baseline method has already surpassed them as they only use the time-domain loss.
In addition, by applying the target-weighted time-frequency loss,
we can obtain better results for both the overlapping regions and the whole audio clip. With the ratio of $\tau$ increasing,
the performance of TSE first raises then decreases, and the best performance can be achieved when $\tau$ is 1.5.
If the ratio is too large, the model will focus too much on the regions that target sound happens, so that the suppression of other regions do not work well. 
\begin{table}[t] \centering
\caption{The influence of mutual learning for TSE and TSD tasks.}
\label{tab:2}
\begin{tabular}{|c|c|c|c|c|}
\hline
\multirow{2}{*}{Stage} & \multicolumn{2}{|c|}{TSE} & \multicolumn{2}{|c|}{TSD} \\
\cline{2-5}
& SI-SDRi & SI-SDRi & segment-F1 & event-F1 \\
\hline
0 & 10.91 & 5.83 & 85.11\% & 43.61\% \\
1 & 11.46 & 5.92 & 85.29\% & 49.26\% \\
2 & 11.67 & 6.10 & 85.93\% & 50.31\% \\
3 & \textbf{11.81} & \textbf{6.14} & \textbf{87.05\%} & \textbf{51.08\%} \\
\hline
\end{tabular}
\end{table}
\begin{figure}[t]
  \centering
  \includegraphics[width=1.0\linewidth]{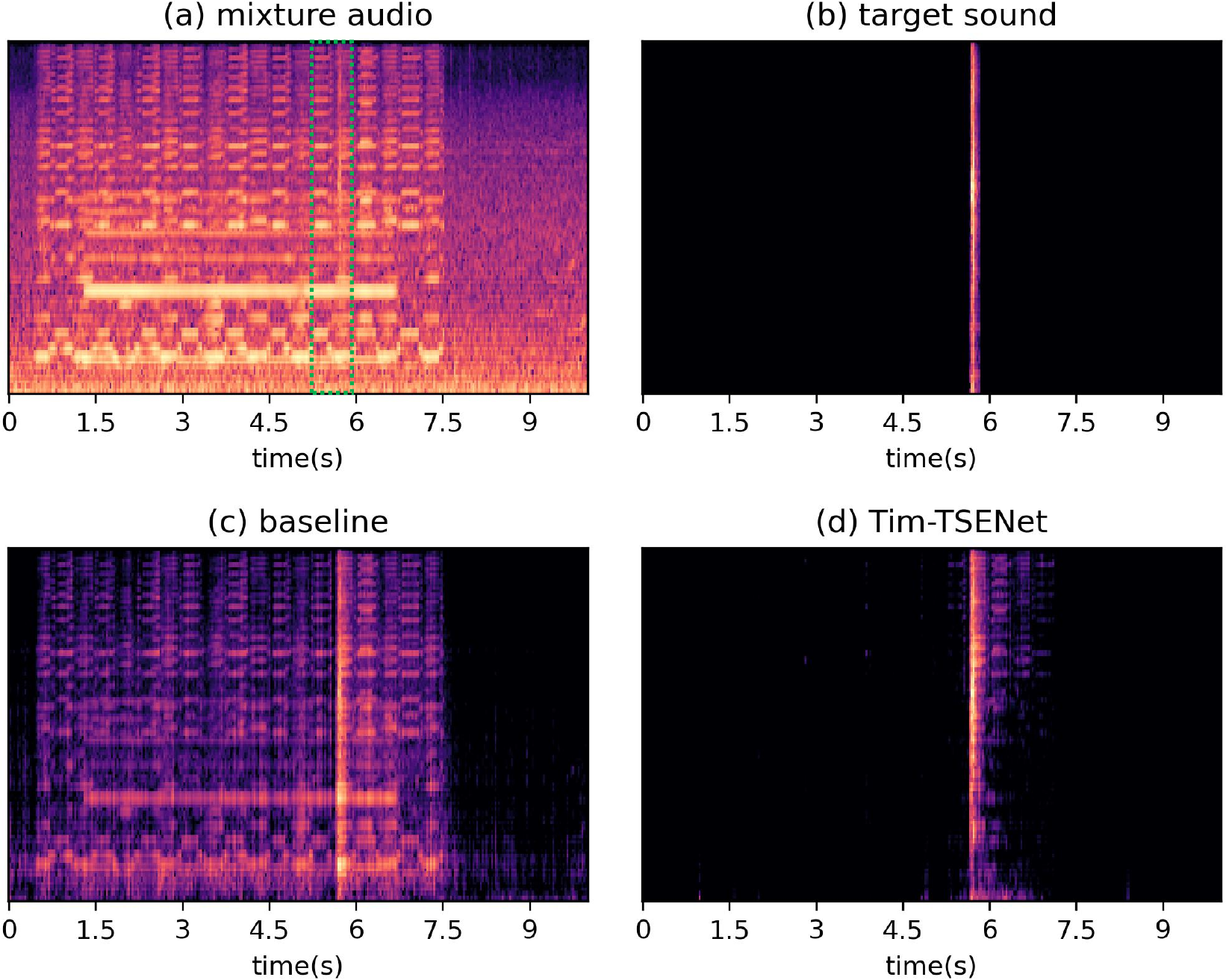}
  \caption{Visualization of the spectrogram of a test example. (a) The input mixture audio. (b) The target sound. (c) The result of our baseline model. (d) The result of Tim-TSENet.}
  \label{fig:2}
\end{figure}
Furthermore, we report the results of three-stage mutual learning framework for TSE and TSD tasks on Table~2.
We find that each stage of mutual learning procedure can lead to the performance gain, especially for the first stage.
After three stages, TSE could get a SI-SDRi of 11.81
and a SI-SDRi-t of 6.14,
while TSD could get a segment-based F score of 87.05\% and an event-based F score of 51.08\%.
Figure~3 gives an example of a test audio, 
where Figure~3(c) shows the spectrogram of the extracted result from our baseline model (without the target-weighted loss and mutual learning) and
Figure~3(d) shows shows the spectrogram of the extracted result from Tim-TSENet. We can see that Tim-TSENet preforms much better than the baseline, especially for the overlapping regions.
More demos can be found here.\footnote{http://dongchaoyang.top/Tim-TSENet-demo/}

\section{Conclusions}
This paper presented a novel network for target sound extraction task, which utilize the timestamp information from a detection network and apply a target-weighted time-frequency loss function to focus more on the target sound. 
We have shown the possibility of mutual learning for an extraction task and a detection task. A future work is to further improve the SI-SDRi-t, and we think designing new sound extractors is a potential solution.
The source code and dateset have been released.\footnote{https://github.com/yangdongchao/Tim-TSENet}

\section{Acknowledgements}
This paper was supported by Shenzhen Science and Technology Fundamental Research Programs JSGG20191129105421211 and GXWD20201231165807007-20200814115301001.

\bibliographystyle{IEEEtran}

\bibliography{mybib}


\end{document}